\documentclass[aps,pra,showpacs,twocolumn]{revtex4-1}
\usepackage{amsmath}
\usepackage{amssymb}
\usepackage{graphicx}
\usepackage{epstopdf}
\usepackage{times,txfonts}
\usepackage{easybmat}
\usepackage{mathtools}
\newcommand{\ket}[1]{|#1\rangle}
\newcommand{\bra}[1]{\langle #1|}
\usepackage[bookmarks=false]{hyperref}
\hypersetup{colorlinks=true,citecolor=blue,linkcolor=blue,urlcolor=blue,pdfstartview=FitH,bookmarksopen=true}
\usepackage{color,soul}

\begin{document}
\title{Approach to realizing nonadiabatic geometric gates with prescribed evolution paths}
\author{K. Z. Li}
\affiliation{Department of Physics, Shandong University, Jinan 250100, China}
\author{P. Z. Zhao}
\email{zhaopeozi2014@foxmail.com}
\affiliation{Department of Physics, Shandong University, Jinan 250100, China}
\author{D. M. Tong}
\email{tdm@sdu.edu.cn}
\affiliation{Department of Physics, Shandong University, Jinan 250100, China}
\date{\today}
\begin{abstract}
Nonadiabatic geometric phases are only dependent on the evolution path of a quantum system but independent of the evolution details, and therefore quantum computation based on nonadiabatic geometric phases is robust against control errors. To realize nonadiabatic geometric quantum computation, it is necessary to ensure that the quantum system undergoes a cyclic evolution and the dynamical phases are removed from the total phases. To satisfy  these conditions, the evolution paths in previous schemes are usually restricted to some special forms, e.g, orange-slice-shaped loops, which make the paths unnecessarily long in general. In this paper, we put forward an approach to the realization of nonadiabatic geometric quantum computation, by which a universal set of nonadiabatic geometric gates can be realized with any desired evolution paths. Our approach makes it possible to realize geometric quantum computation with an economical evolution time so the influence of environment noises on the quantum gates can be minimized further.
\end{abstract}
\maketitle

\section{Introduction}
Quantum computation is believed more effective than classical computation in solving some problems, such as factoring large integers \cite{Shor} and searching unsorted data \cite{Grover}.
The implementation of circuit-based quantum computation relies on a universal set of accurately controllable quantum gates, including arbitrary one-qubit gates and a nontrivial two-qubit gate \cite{Bremner}. However, the inaccurate manipulation of quantum systems inevitably affects the fidelity of quantum gates, and it may seriously spoil the practical realization of quantum computation. To resolve this problem, geometric phases are applied to realizing quantum gates. Since geometric phases are only dependent on the evolution path of a quantum system but independent of the evolution details, quantum gates based on geometric phases are robust against some control errors \cite{Chiara,Solinas2004,Lupo,Filipp,Johansson,Berger,Zhu2005}.

The first scheme of geometric quantum computation \cite{Jones} is based on adiabatic Abelian geometric phases, i.e., Berry phases \cite{Berry}, and the adiabatic geometric quantum computation was soon extended to adiabatic holonomic quantum computation \cite{Duan,Zanardi} based on adiabatic non-Abelian geometric phases \cite{Wilczek}. Since adiabatic geometric or holonomic quantum computation requires the quantum system to evolve slowly enough, it needs a long run time to perform geometric gates. However, a long run time can make the quantum gate vulnerable to the environment-induced decoherence. To relax the limit of evolution speed, nonadiabatic geometric quantum computation \cite{WangXB,Zhu} based on Aharonov-Anandan phases \cite{Aharonov} and nonadiabatic holonomic quantum computation \cite{Sjoqvist2012,Xu2012} based on nonadiabatic non-Abelian geometric phases \cite{Anandan} were then proposed. Nonadiabatic geometric quantum computation can be realized with a high-speed implementation. Due to the merits of both geometric robustness and high-speed evolution, nonadiabatic geometric quantum computation has attracted much attention \cite{Zhu2003,Zheng,Sjoqvist2003,Chen2006,Cen2006,Feng2007,Kim,Wu2007,
Feng2009,Chen2012,Zhao2016,Zhu2003PRA,Ota,Zhang2005,Solinas,Oto2009,Thomas2011,Xu2014PRA,Xu2014SR,Zhao,Xue,Liu,
Du,Leibfried,Yin,Sun}, and has been experimentally demonstrated with trapped ions \cite{Leibfried}, nuclear magnetic resonance \cite{Du} and superconducting circuits \cite{Yin,Sun}.

To realize nonadiabatic geometric quantum computation, it is necessary to ensure that the quantum system undergoes a cyclic evolution and the dynamical phases are removed from the total phases. To satisfy  these requirements, the evolution paths in the previous works were mainly restricted to some special forms such as early multiple loops \cite{Zhu,Zhu2003PRA,Ota} and the widely used orange-slice-shaped loops\cite{Solinas,Oto2009,Thomas2011,Xu2014PRA,Xu2014SR,Zhao,Xue}.  The multiple-loop scheme makes the quantum system evolve along several closed loops such that the dynamical phases in different loops cancel each other. The orange-slice-shaped-loop scheme makes the quantum system evolve along the geodesics on the Bloch sphere such that the dynamical phases are always zero during the evolution. Although the paths in the orange-slice-shaped-loop scheme are generally shorter than those in the multiple-loop scheme, they are still unnecessarily lengthy for realizing most of the geometric gates, especially for the gates with small rotation angles. Roughly speaking, a long evolution path corresponds to a long evolution time, and therefore implies a long exposure of the system to environment noises. Hence, it is an interesting topic to optimize the evolution paths of the quantum system \cite{Glaser} for realizing nonadiabatic geometric quantum computation. In this paper, we put forward an approach for the realization of nonadiabatic geometric quantum computation, by which a universal set of nonadiabatic geometric gates can be realized  with any desired evolution paths. Our approach allows us to realize geometric gates with shorter evolution paths than those in the previous schemes and makes it possible to minimize the evolution time so the influence of environment noises on the quantum gates can be reduced.

\section{General form of Hamiltonian}

In this section, we give a general form of Hamiltonians that can be used to realize nonadiabatic geometric quantum computation. To this end, we first recall the notions of nonadiabatic geometric phases and the quantum computation based on them. Consider an $N$-dimensional quantum system defined by Hamiltonian $H(t)$. Its time-dependent quantum state is denoted as $\ket{\phi(t)}$. If the quantum system evolves cyclically in the state space with period $\tau$, i.e., $\ket{\phi(\tau)}=\exp[i\alpha(\tau)]\ket{\phi(0)}$, then the total phase $\alpha(\tau)$, accumulated during the cyclic evolution, can be expressed as $\alpha(\tau)= \beta(\tau) + \gamma(\tau)$, where $\beta(\tau)=-\int^{\tau}_{0}\bra{\phi(t)}H(t)\ket{\phi(t)}dt$ is the dynamical phase and $\gamma(\tau):=\alpha(\tau)- \beta(\tau)$ is known as the nonadiabatic geometric phase \cite{Aharonov}.
To realize a universal set of geometric gates, we need to encode one or two logical qubits into a two-dimensional or four-dimensional subspace of an $N$-dimensional quantum system consisting of a number of physical qubits, and make the states in the subspace evolve along some special paths that correspond to a zero dynamical phase.

With these notions, we may now consider an inverse problem. We ask how to find a Hamiltonian $H(t)$ which can drive a quantum system from the initial state $\ket{\phi_k(0)}$ to the final state $\ket{\phi_k(\tau)}$  along a desired path prescribed by $\ket{\phi_k(t)}$, such that
\begin{align}
\ket{\phi_k(\tau)}=e^{i\gamma_k(\tau)}\ket{\phi_k(0)},\label{Tong1}
\end{align}
and
\begin{align}
\bra{\phi_k(t)}H(t)\ket{\phi_k(t)}=0.\label{Tong2}
\end{align}
Here, $\ket{\phi_k(t)}$, $k=1,2,\cdots, N$, satisfy $\bra{\phi_i(0)}\phi_j(0)\rangle =\delta_{ij}$ and  $\gamma_k(\tau)$ are the geometric phases.  Expressions (\ref{Tong1})  and (\ref{Tong2}) are often known as the cyclic evolution condition and the parallel transport condition, respectively. As $\ket{\phi_k(t)}$ satisfies the Schr\"{o}dinger equation, $i\ket{\dot{\phi}_k(t)}=H(t)\ket{\phi_k(t)}$, the parallel transport condition can be equivalently written as $\bra{\phi_k(t)}\dot{\phi}_k(t)\rangle=0$.

To find such Hamiltonians,  we introduce a set of auxiliary orthonormal vectors $\{\ket{\nu_k(t)}\}^{N}_{k=1}$ defined by $\ket{\nu_{k}(t)}=\exp[-i\gamma_k(t)]\ket{\phi_{k}(t)}$, where $\gamma_k(t)$ are the real functions that make $\ket{\nu_{k}(\tau)}=\ket{\nu_{k}(0)}=\ket{\phi_{k}(0)}$, i.e., $\gamma_k(t)|_{t=0}=0$ and  $\gamma_k(t)|_{t=\tau}=\gamma_k(\tau)$. By requiring $\ket{\phi_{k}(t)}=\exp[i\gamma_k(t)]\ket{\nu_{k}(t)}$ to satisfy Eq. (\ref{Tong2}), we immediately have
\begin{align}\label{eq}
\gamma_k(t)=i\int^{t}_{0}\langle\nu_k(t^{\prime})\ket{\dot{\nu}_k(t^{\prime})}dt^{\prime}.
\end{align}
Substituting $\ket{\phi_{k}(t)}=\exp[-\int^{t}_{0}\langle\nu_k(t^{\prime})\ket{\dot{\nu}_k(t^{\prime})}dt^{\prime}]\ket{\nu_{k}(t)}$ into $H(t)=i\sum^{N}_{k=1}\ket{\dot{\phi}_k(t)}\bra{\phi_k(t)}$, we obtain
\begin{align}\label{eq1}
H(t)=i\sum_{l\neq k}^{N}\langle\nu_{l}(t)\ket{\dot{\nu}_{k}(t)}\ket{\nu_{l}(t)}\bra{\nu_{k}(t)}.
\end{align}

It is easy to verify that $\ket{\phi_k(t)}$, $k=1,2,\cdots, N$, satisfy the Schr\"{o}dinger equation $i\ket{\dot{\phi}_k(t)}=H(t)\ket{\phi_k(t)}$ as well as the cyclic evolution condition (\ref{Tong1}) and parallel transport condition (\ref{Tong2}). Therefore, if the quantum system governed by the Hamiltonian in Eq. (\ref{eq1}) is initially in the state  $\ket{\phi_k(0)}$, it will evolve along the path given by  $\ket{\phi_k(t)}$, and arrive at the final state $\ket{\phi_k(\tau)}=\exp[i\gamma_k(\tau)]\ket{\phi_k(0)}$ with $\gamma_k(\tau)$ being a purely geometric phase.  Consequently, the evolution operator at time $t=\tau$ reads
\begin{align}\label{eq2}
U(\tau)=\sum_{k}e^{i\gamma_k(\tau)}\ket{\phi_{k}(0)}\bra{\phi_{k}(0)}.
\end{align}
We can use the Hamiltonians  to realize nonadiabatic geometric quantum computation by encoding logical qubits into a subspace of $\text{span}\{ \ket{\phi_1(0)}, \ket{\phi_2(0)}, \cdots,\ket{\phi_N(0)}\}$. It is worth noting that it is not necessary to require all the solutions $\ket{\phi_{k}(t)}$ satisfy the conditions Eqs. (\ref{Tong1}) and (\ref{Tong2}). In fact, to realize a universal set of geometric gates, it is sufficient to have a $L$-dimensional subspace with $L=2$ for one-qubit gates or $L=4$ for two-qubit gates.

\section{Implementation}

The above discussion shows that starting from an arbitrary set of auxiliary bases $\{\ket{\nu_{1}(t)}, \ket{\nu_{2}(t)},\cdots, \ket{\nu_{N}(t)}\}$ with $\ket{\nu_k(\tau)}= \ket{\nu_k(0)}$, we can easily write out the Hamiltonian by using Eq. (\ref{eq1}) with $\{\ket{\phi_1(t)}, \ket{\phi_2(t)}\cdots,\ket{\phi_{N}(t)}\}$, defined by $\ket{\phi_{k}(t)}=\exp[-\int^{t}_{0}\langle\nu_k(t^{\prime})\ket{\dot{\nu}_k(t^{\prime})}dt^{\prime}]\ket{\nu_{k}(t)}$, being the solutions of the Schr\"{o}dinger equation. The space $\text{span}\{\ket{\phi_1(t),\phi_2(t),\cdots, \phi_L(t)}\}$ or any subspace of it naturally satisfies both the cyclic evolution and parallel transport conditions, and therefore can be taken as the computational space of nonadiabatic geometric computation. As a result, the nonadiabatic geometric gate in Eq. (\ref{eq2}) can be realized. Due to the arbitrariness of auxiliary bases $\{\ket{\nu_1(t)}, \ket{\nu_2(t)},\cdots, \ket{\nu_{N}(t)}\}$, our approach for constructing the Hamiltonian allows quantum systems to be driven along any desired evolution paths and thus makes it possible to minimize the evolution time needed for realizing geometric gates. To show its usefulness, we will give a universal set of nonadiabatic geometric gates, i.e., arbitrary one-qubit gates and a non-trivial two-qubit gate, from which one will see that the Hamiltonians used in the previous orange-slice-shaped-loop schemes of nonadiabatic geometric quantum computation are only special cases of the general formalism, and an alternative choice of the Hamiltonian can markedly reduce the evolution time.

\subsection{One-qubit gate}

To realize nonadiabatic geometric gates, we only need to consider a two-level quantum system with the bare states $\ket{0}$ and $\ket{1}$. We choose the auxiliary bases as
\begin{align}\label{T1}
&\ket{\nu_{1}(t)}=\cos\frac{\theta(t)}{2}\ket{0}+\sin\frac{\theta(t)}{2}e^{i\varphi(t)}\ket{1},
\notag\\
&\ket{\nu_{2}(t)}=\sin\frac{\theta(t)}{2}e^{-i\varphi(t)}\ket{0}-\cos\frac{\theta(t)}{2}\ket{1},
\end{align}
where $\theta(t)$ and $\varphi(t)$ are the time-dependent parameters.
In this case, from Eq. (\ref{eq1}), we have the Hamiltonian,
\begin{align}\label{eq3}
H(t)=&-\frac{1}{2}\Big[\dot{\theta}(t)\sin\varphi(t)
+\dot{\varphi}(t)\sin\theta(t)\cos\theta(t)\cos\varphi(t)\Big]\sigma_{x}
\notag\\
&+\frac{1}{2}\Big[\dot{\theta}(t)\cos\varphi(t)
-\dot{\varphi}(t)\sin\theta(t)\cos\theta(t)\sin\varphi(t)\Big]\sigma_{y}
\notag\\
&+\frac{1}{2}\dot{\varphi}(t)\sin^{2}\theta(t)\sigma_{z},
\end{align}
where $\sigma_{x}$, $\sigma_{y}$ and $\sigma_{z}$ are the standard Pauli operators acting on $\ket{0}$ and $\ket{1}$.
This Hamiltonian describes a general two-level system driven by a near resonate laser or microwave, of which the detuning and Rabi frequency are corresponding to $\Delta(t)=-\dot{\varphi}(t)\sin^{2}\theta(t)/2$ and  $\Omega(t)=-\dot{\theta}(t)\exp\{i[\varphi(t)-\pi/2]\}/2-\dot{\varphi}(t)\sin\theta(t)\cos\theta(t)\exp[-i\varphi(t)]/2$, respectively. With the control field parameters $\Delta(t)$ and $\Omega(t)$, it can be briefly written as $H(t)=\Delta(t)(\ket{1}\bra{1}-\ket{0}\bra{0})+[\Omega(t)\ket{1}\bra{0}+\mathrm{H.c.}]$.  Such a Hamiltonian can be realized in many physical systems, such as trapped ions, nuclear magnetic resonance, superconducting circuits, nitrogen-vacancy centers in diamond, and Rydberg atoms.

Corresponding to the auxiliary bases given in Eqs. (\ref{T1}), the initial states $\ket{\phi_1(0)}$ and $\ket{\phi_2(0)}$  are
\begin{align}\label{eq4}
\ket{\phi_1(0)}&=\ket{\nu_{1}(0)}=\cos\frac{\theta_0}{2}\ket{0}+\sin\frac{\theta_0}{2}e^{i\varphi_0}\ket{1},
\notag\\
\ket{\phi_{2}(0)}&=\ket{\nu_{2}(0)}=\sin\frac{\theta_0}{2}e^{-i\varphi_0}\ket{0}-\cos\frac{\theta_0}{2}\ket{1},
\end{align}
where $\theta_0=\theta(0)$ and $\varphi_0=\varphi(0)$. Starting from a state $\{\ket{\phi_1(0)}$ or $\ket{\phi_2(0)}\}$, the quantum system driven by the Hamiltonian will undergo a cyclic evolution. The unitary operator at the final time $\tau$ is given by Eq. (\ref{eq2}). It reads
\begin{align}\label{eq6}
U(\tau)=e^{-i\gamma(\tau)}\ket{\phi_{1}(0)}\bra{\phi_{1}(0)}
+e^{i\gamma(\tau)}\ket{\phi_{2}(0)}\bra{\phi_{2}(0)}
\end{align}
with
\begin{align}\label{eq5}
 \gamma(\tau)=\frac{1}{2}\int^{\tau}_{0}[1-\cos\theta(t)]\dot{\varphi}(t)dt.
\end{align}
It can be further expressed as
\begin{align}\label{Tong3}
U(\tau)=e^{-i\gamma(\tau)\boldsymbol{\mathrm{n}\cdot\sigma}},
\end{align}
where $\boldsymbol{\mathrm{n}}=(\sin\theta_{0}\cos\varphi_{0},\sin\theta_{0}\sin\varphi_{0},\cos\theta_{0})$ is a unit vector and $\boldsymbol{\sigma}=(\sigma_{x},\sigma_{y},\sigma_{z})$ are Pauli operators.
Obviously, $U(\tau)$ represents an arbitrary one-qubit gate with the rotation axis $\boldsymbol{\mathrm{n}}$ and rotation angle $2\gamma(\tau)$.

It is interesting to see that $\gamma(\tau)$ can be related with a solid angle in the parameter space determined by $\theta(t)$ and $\varphi(t)$. If we take $\theta(t)$ as the polar angle and $\varphi(t)$ as azimuthal angle of a spherical coordinate system, then $(\theta(t), ~\varphi(t))$ represents a point in a unit two-sphere, and it traces a closed path $C$ in the unit sphere as  the quantum system evolves from $t=0$ to $t=\tau$. From this perspective, Eq. (\ref{eq5}) can be recast as
\begin{align}\label{eq7}
 \gamma(\tau)=\frac{1}{2}\oint_{C}(1-\cos\theta)d\varphi.
\end{align}
It just represents a half of the solid angle enclosed by path $C$ in the unit sphere. This implies that $\gamma(\tau)$ is only dependent on the evolution path traced by the parameters $\theta(t)$ and $\varphi(t)$ but independent of the evolution details such as the changing rate of the parameters, being robust against the control errors that do not change the evolution path but affect other aspects of the evolution. In fact, $\gamma(\tau)$ is still invariant even if the path is changed but as long as the area enclosed by the path is unchanged.

The above discussion indicates that the rotation axis is only dependent on the initial values of the parameters $\theta_{0}$ and $\varphi_{0}$ and the phase value $\gamma(\tau)$ is only dependent on the solid angle enclosed by the evolution path. To perform a nonadiabatic geometric gate with rotation axis $\boldsymbol{\mathrm{n}}$ and rotation angle $2\gamma(\tau)$, infinitely many evolution paths with different lengths can be chosen. Some of them are shorter than others and a shorter path implies a shorter evolution time in general. By using our approach, one can choose any desired evolution path to realize a given geometric gate.
In the following, we take $U(\tau)=\exp{(-i\pi\sigma_{z}/8)}$, of which $\boldsymbol{\mathrm{n}}=(0,0,1)$ and $\gamma(\tau)=\pi/8$, as an example to illustrate this point.

\begin{figure}[t]
\begin{center}
\includegraphics[scale=0.32]{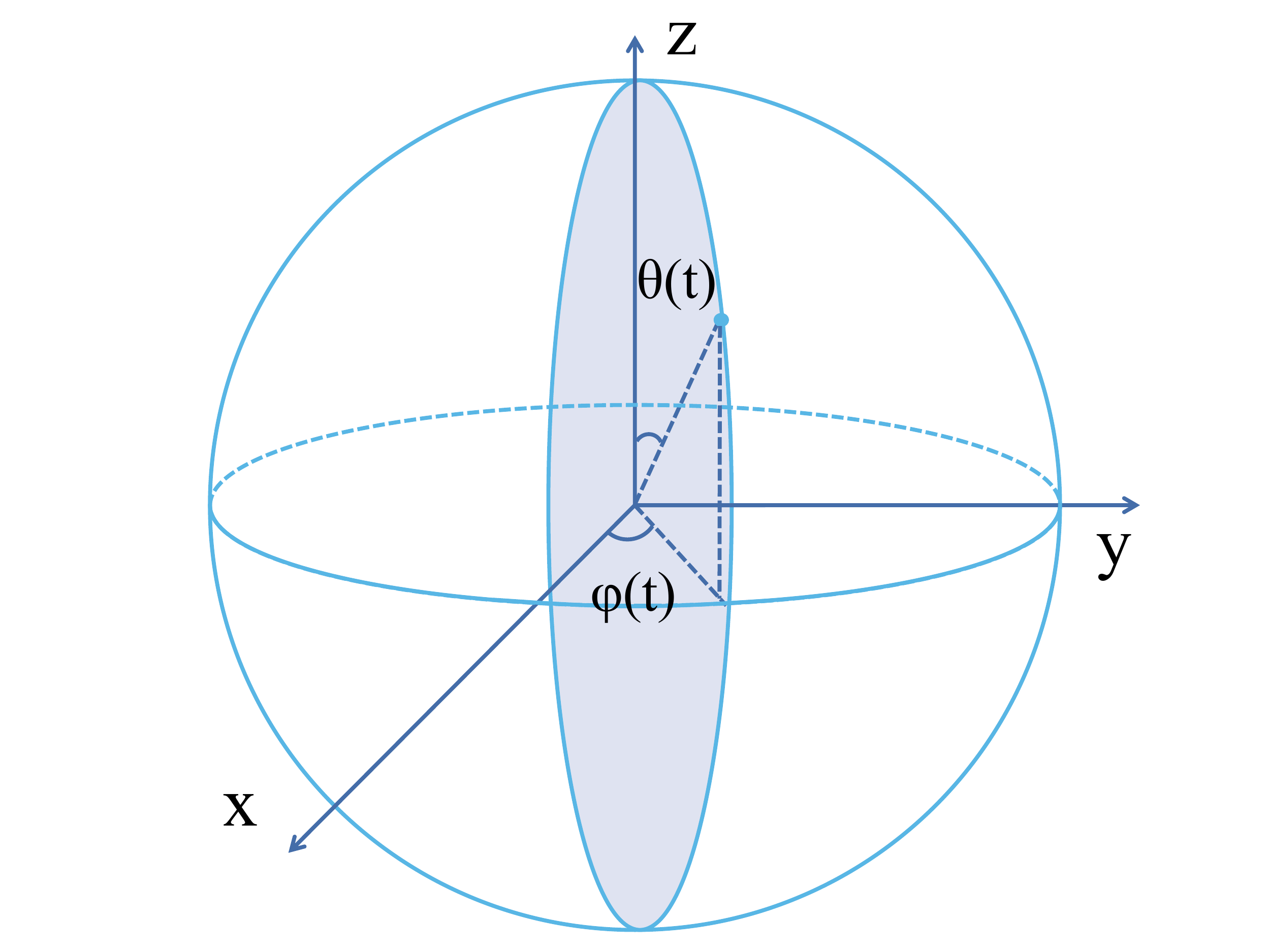}
\end{center}
\caption{(Color online) The orange-slice-shaped-loop path for the realization $U(\tau)=\exp{(-i\pi\sigma_{z}/8)}$.   \label{Fig1}}
\end{figure}

To realize $U(\tau)=\exp{(-i\pi\sigma_{z}/8)}$, we can choose the evolution path shown in Fig. \ref{Fig1}. The parameters $(\theta(t),\varphi(t))$ start from the north pole $(0,\varphi_{0})$ to the south pole $(\pi,\varphi_{0})$ along the great circle $\varphi(t)=\varphi_{0}$, then return back to the north pole from the south pole along another great circle  $\varphi(t)=\varphi_{0}+\pi/8$.
For this path, the Hamiltonian can be obtained by using Eq. (\ref{eq3}). It reads
$H(t)=\Omega(t)(-\sin\varphi_{0}\sigma_{x}+\cos\varphi_{0}\sigma_{y})$ for $t\in[0,\tau/2]$ and $H(t)=\Omega(t)[-\sin(\varphi_{0}+\pi/8)\sigma_{x}
+\cos(\varphi_{0}+\pi/8)\sigma_{y}]$ for $t\in(\tau/2,\tau]$. Here, $\Omega(t)=\dot{\theta}(t)/2$  plays the role of laser pulse envelope satisfying
$\int^{\tau/2}_{0}\Omega(t)dt=\pi/2$ and $\int^{\tau}_{\tau/2}\Omega(t)dt=-\pi/2$.
The evolution time is about $\tau\sim\pi/\bar{\Omega}$ with $\bar{\Omega}$ being the average modulus of $\Omega(t)$, and the length of the evolution path is exactly $2\pi$. This path is just the orange-slice-shaped loop widely used in the previous schemes of nonadiabatic geometric quantum computation \cite{Solinas,Oto2009,Thomas2011,Xu2014SR,Xu2014PRA,Zhao}.

\begin{figure}[t]
\begin{center}
\includegraphics[scale=0.32]{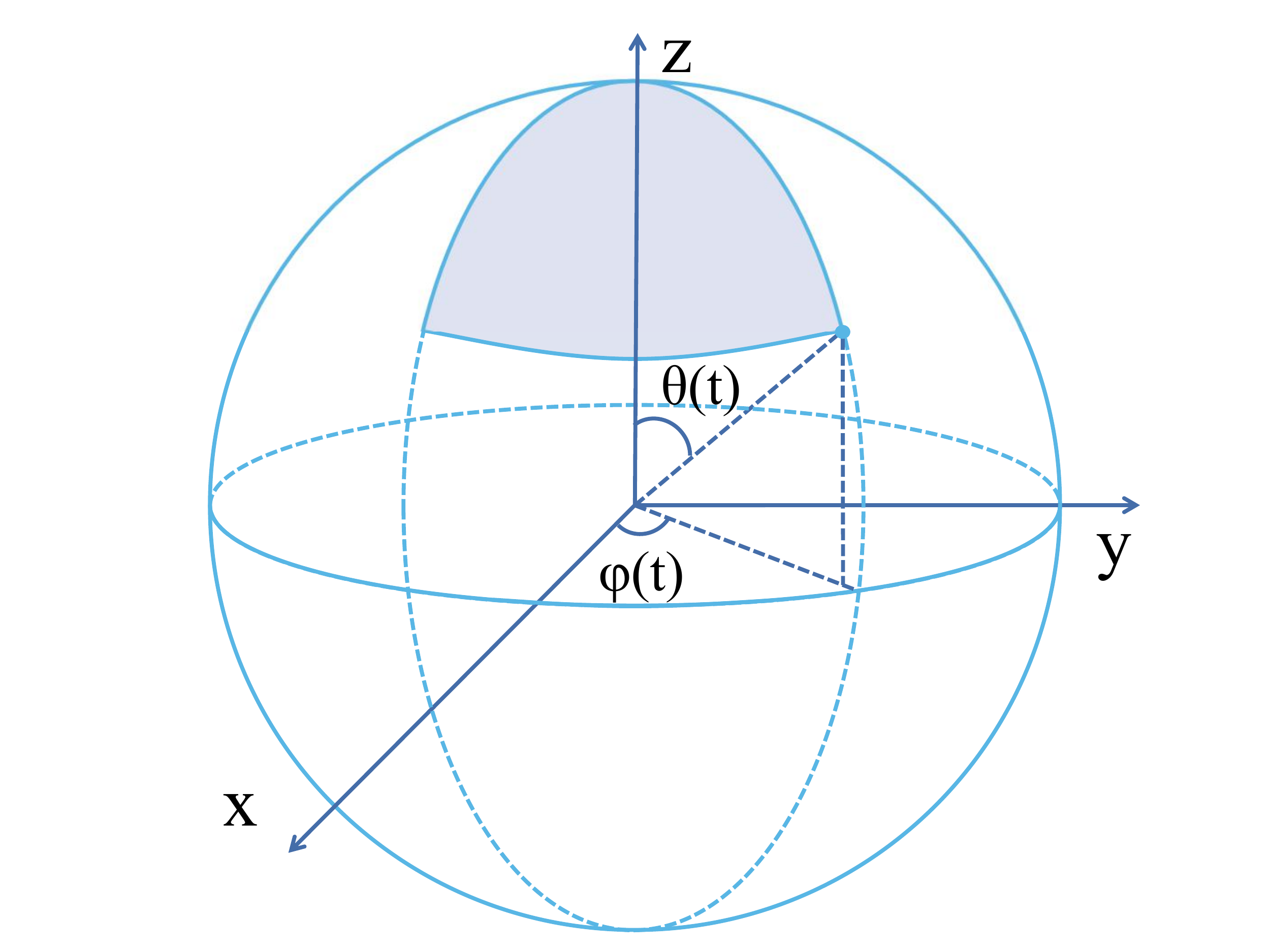}
\end{center}
\caption{(Color online) The optimized path for the realization $U(\tau)=\exp{(-i\pi\sigma_{z}/8)}$.   \label{Fig2}}
\end{figure}

To realize this gate, we can also choose the alternative evolution path shown in Fig. \ref{Fig2}. The parameters $(\theta(t),\varphi(t))$ start from the north pole $(0,\varphi_{0})$ to the point  $(\pi/3,\varphi_{0})$ along the great circle $\varphi(t)=\varphi_{0}$, then evolve from $(\pi/3,\varphi_{0})$  to $(\pi/3,\varphi_{0}+\pi/2)$ along the arc $\theta(t)=\pi/3$, and finally return back to the north pole from the point $(\pi/3,\varphi_{0}+\pi/2)$ along the great circle $\varphi(t)=\varphi_{0}+\pi/2$.
For this path, the piecewise Hamiltonian reads $H(t)=\Omega(t)(-\sin\varphi_{0}\sigma_{x}+\cos\varphi_{0}\sigma_{y})$ for $t\in[0,\tau_{1}]$, $H(t)=\tilde{\Omega}(t)[\cos\varphi(t)\sigma_{x}+\sin\varphi(t)\sigma_{y}]+\Delta(t)\sigma_{z}$ for  $t\in(\tau_{1},\tau_{2}]$, and $H(t)=-\Omega(t)(\cos\varphi_{0}\sigma_{x}
+\sin\varphi_{0}\sigma_{y})$ for $t\in(\tau_{2},\tau]$, where $\Omega(t)=\dot{\theta}(t)/2$,  $\tilde{\Omega}(t)=-\sqrt{3}\dot{\varphi}(t)/8$ and $\Delta(t)=3\dot{\varphi}(t)/8$. Here, $\Omega(t)$ and $\tilde{\Omega}(t)$ are the laser pulse envelopes satisfying
$\int^{\tau_{1}}_{0}\Omega(t)dt=\pi/6$, $\int^{\tau_{2}}_{\tau_{1}}\tilde{\Omega}(t)dt=-\sqrt{3}\pi/16$, and
$\int^{\tau}_{\tau_{2}}\Omega(t)dt=-\pi/6$, and  $\Delta(t)=3\dot{\varphi}(t)/8$ is the laser pulse detuning. At a rough estimate, the evolution time is about $\tau\sim0.44\pi/\bar{\Omega}$ with $\bar{\Omega}$ being the average modulus of these laser pulse envelopes. The length of this path is $7\pi/6$, which is shorter than that of the orange-slice-shaped loop.

Obviously, we can choose a much shorter evolution path to realize the same gate. Since the rotation axis of this gate is corresponding to the north pole and the circumference of a circle is shorter than that of any other shapes in the case of the same solid angel enclosed, the shortest evolution path is the circles that start from the north pole and enclose a solid angle of $\pi/4$.  The length of the shortest path is $\sqrt{15}\pi/4$. Certainly, this point is valid for the general case with $U(\tau)=\exp[-i\gamma(\tau)\boldsymbol{\mathrm{n}\cdot\sigma}]$, for which the shortest evolution path is the circles that start from the point $(\theta_{0},~\varphi_{0})$ and enclose a solid angle of $2\gamma(\tau)$.
Note that $\varphi(\tau)$ is not required to be equal to $\varphi(0)$ in the special case of $\theta(0)=0$ or $\pi$ but it is necessary to have  $\varphi(\tau)=\varphi(0)$ in all other cases.

\subsection{Two-qubit gate}

To perform nonadiabatic  geometric quantum computation, a nontrivial two-qubit gate is needed besides the arbitrary one-qubit gates given above. We now demonstrate how to realize a nontrivial two-qubit gate by using our approach. To this end, we choose the auxiliary bases as
\begin{align}
\ket{\nu_{1}(t)}=&\ket{00},\notag\\
\ket{\nu_{2}(t)}=&\ket{11},
\notag\\
\ket{\nu_{3}(t)}=&\cos\frac{\alpha(t)}{2}\ket{01}+\sin\frac{\alpha(t)}{2}e^{i\beta(t)}\ket{10},
\notag\\
\ket{\nu_{4}(t)}=&\sin\frac{\alpha(t)}{2}e^{-i\beta(t)}\ket{01}-\cos\frac{\alpha(t)}{2}\ket{10},
\end{align}
where $\alpha(t)$ and $\beta(t)$ are time-dependent parameters.
By using Eq. (\ref{eq1}), we can easily write out the Hamiltonian,
\begin{align}\label{ha}
H(t)=&-\frac{1}{2}\Big[\dot{\alpha}(t)\sin\beta(t)
+\dot{\beta}(t)\sin\alpha(t)\cos\alpha(t)\cos\beta(t)\Big]R^{x}
\notag\\
&-\frac{1}{2}\Big[\dot{\alpha}(t)\cos\beta(t)
-\dot{\beta}(t)\sin\alpha(t)\cos\alpha(t)\sin\beta(t)\Big]R^{y}
\notag\\
&+\frac{1}{2}\dot{\beta}(t)\sin^{2}\alpha(t)R^{z},
\end{align}
where $R^{x}=(\sigma_{x}\sigma_{x}+\sigma_{y}\sigma_{y})/2$ is the $XY$ interaction, $R^{y}=(\sigma_{x}\sigma_{y}-\sigma_{y}\sigma_{x})/2$ is the Dzialoshinski-Moriya interaction, and $R^{z}=(\sigma^{1}_{z}-\sigma^{2}_{z})/2$ relates to the local Pauli operators acting on single qubits.

This Hamiltonian is experimentally feasible. For example, it can be realized in the system of trapped ions with the S{\o}rensen$-$M{\o}lmer setting \cite{SM,SM2000}.  Specifically, we consider two two-level trapped ions. We use a blue sideband laser with  Rabi frequency $\Omega_{1}(t)$ and detuning $-(\nu+\delta)$  to drive the transitions $\ket{0}\leftrightarrow\ket{1}$  of the first ion, and use another blue sideband laser with Rabi frequency $\Omega_{2}(t)$ and the same detuning $-(\nu+\delta)$  to drive the transitions $\ket{0}\leftrightarrow\ket{1}$  of the second ion, where $\nu$ is the frequency of the vibrational mode of the trapped ions and $\delta$ is an additional detuning. In the rotating frame and with the rotating-wave approximation, the Hamiltonian of the two-ion system reads
\begin{align}\label{eq8}
H(t)=&i\eta\Omega_{1}(t)e^{-i\delta t}a^{\dagger}\ket{1}_{11}\bra{0}
\notag\\
&+i\eta\Omega_{2}(t)e^{-i\delta t}a^{\dagger}\ket{1}_{22}\bra{0}+\mathrm{H.c.}
\end{align}
in the Lamb-Dicke regime. Here, $\ket{\cdot}_{j}$ represents the states of the $j$th ion, $a$ and $a^{\dagger}$ are the annihilation and creation operators of the vibrational mode, respectively, and $\eta$ is the
Lamb-Dicke parameter, which satisfies $\eta^{2}(n+1)\ll1$ with $n$ being the quantum number of the vibrational mode. If the large detuning condition $\delta\gg\eta\Omega_{1}(t),\eta\Omega_{2}(t)$ is satisfied, the Hamiltonian is reduced to an effective one \cite{eff},
\begin{align}\label{eq9}
H(t)=\Omega_{\mathrm{eff}}(t)\ket{01}\bra{10}+\mathrm{H.c.},
\end{align}
where $\Omega_{\mathrm{eff}}(t)=\eta^{2}\Omega^{\ast}_{1}(t)\Omega_{2}(t)/\delta$. The effective Hamiltonian describes the $XY$ interaction if $\Omega_{\mathrm{eff}}(t)$ is a real number, and it describes the Dzialoshinski-Moriya interaction if  $\Omega_{\mathrm{eff}}(t)$ is an imaginary number. Therefore, the $XY$ interaction as well as the Dzialoshinski-Moriya interaction can be realized with the tapped ions driven by lasers. As for the local Pauli operators, they can be easily realized by using a large detuning laser acting on each of the ions. In summary, the Hamiltonian in Eq. (\ref{ha}) is  experimentally feasible.

Starting from a state in the space $\{\ket{\phi_k(0)}\}^{4}_{k=1}$ with $\ket{\phi_k(0)}=\ket{\nu_{1}(0)}$, the quantum system driven by the Hamiltonian will undergo a cyclic evolution.  According to Eq. (\ref{eq2}), we obtain
\begin{align}
U(\tau)=&\ket{00}\bra{00}+\ket{11}\bra{11}+e^{-i\gamma(\tau)}\ket{\phi_{3}(0)}\bra{\phi_{3}(0)}
\notag\\
&+e^{i\gamma(\tau)}\ket{\phi_{4}(0)}\bra{\phi_{4}(0)},
\end{align}
where $\gamma(\tau)=\int^{\tau}_{0}[1-\cos\alpha(t)]d\beta(t)/2$. The unitary operator $U(\tau)$ is a nontrivial two-qubit geometric gate. Since the expression of $\gamma(\tau)$ is similar to Eq. (\ref{eq5}), we can have further discussions on the two-qubit gate, as done on one-qubit gates, and can find all the interesting properties similar to the one-qubit case. For example, $\gamma(\tau)$ can be expressed as $\gamma(\tau)=\frac{1}{2}\oint_{C}(1-\cos\alpha)d\beta$ in the parameter space  determined by $\alpha(t)$ and $\beta(t)$, and is robust against the control errors that do not change the area enclosed by the evolution path.

\section{Conclusion}
In conclusion, we have proposed an approach for the realization of nonadiabatic geometric quantum computation. Our result shows that starting from a set of auxiliary bases $\ket{\nu_k(t)}$ , $k=1,2,\dots,N$, with $\ket{\nu_k(\tau)}=\ket{\nu_k(0)}$, one can easily write out the Hamiltonian $H(t)=i\sum_{l\neq k}^{N}\langle\nu_{l}(t)\ket{\dot{\nu}_{k}(t)}\ket{\nu_{l}(t)}\bra{\nu_{k}(t)}$, which can make the quantum system evolve from  initial state $\ket{\phi_k(0)}=\ket{\nu_k(0)}$  to final state $\ket{\phi_k(\tau)}=\exp[i\gamma(\tau)]\ket{\phi_k(0)}$ with $\gamma(\tau)=i\int^{\tau}_{0}\langle\nu_k(t)\ket{\dot{\nu}_k(t)}dt$ being a purely geometric phase.
By encoding logical qubits into a subspace spanned by $\{\ket{\nu_1(0)}, \ket{\nu_2(0)}, \cdots,\ket{\nu_N(0)}\}$, we can use the Hamiltonians  to realize nonadiabatic geometric gates with any desired evolution paths since the choices of $\{\ket{\nu_k(t)}\}^{N}_{k=1}$ are flexible. Our approach makes it possible to minimize the evolution time needed for realizing a geometric gate. To show its application, we have given a universal set of geometric gates with much shorter evolution time than that in the previous schemes.

\begin{acknowledgments}
K.Z.L. acknowledges support from the National Natural Science Foundation of China through Grant No. 11775129. P.Z.Z. acknowledges support from the China Postdoctoral Science Foundation though Grant No. 2019M662318. D.M.T. acknowledges support from the National Natural Science Foundation of China through Grant No. 11575101.
\end{acknowledgments}


\begin{thebibliography}{99}
\bibitem{eff} By following the approach in Ref. \cite{James}, the Hamiltonian in Eq. (\ref{eq8}) can be reduced to $H(t)=\Omega_{\mathrm{eff}}(t)\ket{01}\bra{10}+\mathrm{H.c.}+\mathrm{Stark~shift~terms}$.
    The Stark shift terms can be easily compensated by using additional lasers \cite{Haffner}, and we finally obtain the effective Hamiltonian in Eq. (\ref{eq9}).
\end{thebibliography}
\end{document}